# Open Source Software (OSS) Transparency for DoD Acquisition

## Abstract

Caveat emptor, or "let the buyer beware," is commonly attributed to open source software (OSS)—the onus is on the OSS consumer to ensure that it is fit for use in the consumer's context. OSS has been compared to an open market bazaar where consumers are free to browse all the source code and take a copy. But there are a few fundamental problems with such an analogy:

- The consumer must have the wherewithal and skills to comprehend the source code in a manner that allows them to use it effectively, which might exceed the skills of the myriad authors who produced that software in the first place.
- The consumer also lacks the insight into the practices exercised by the authors in the production of that source. Such practices include code quality checks, peer reviews, software testing, and secure software development practices.

The burden on the consumer is considerable. Consumers—both individuals and organizations—have access to proprietary and open source tools to help them analyze source code as a means of understanding what is good and what is problematic if they are aware of and use those tools.

What a consumer lacks is an understanding of how OSS differs from other third-party software products. OSS contributors and maintainers may be arbiters for the OSS product, but they are not selling it. Caveat emptor, therefore, applies not only to the OSS product the consumer wants to use, but it also applies to the consumer knowing more about the OSS project, its contributors and maintainers, as well as the processes they follow. To gain that insight, transparency is needed for consumers to have access to the data and information about the OSS project that is meaningful from a software integrity and assurance perspective and suitable for supply chain risk analysis. The Open Source Security Foundation's (OSSF's) SLSA2 and CHAOSS3 are initial steps in this space. The Software Engineering Institute (SEI), MITRE Corporation, and the Carnegie Mellon University Open Source Programs Office (CMU OSPO) are currently working on effective measures for supply chain risk management as it pertains to OSS. This collaborative effort is proposing forming a special interest group within the OSSF. This paper shares the results to date in this area of research.

# Introduction

## Scope of the Open Source Software (OSS) Problem Space

By some counts, there were over 53 million projects in 2022 alone at one OSS repository: GitHub (Woodward, 2022). Software from repositories and sources likes these are included in practically every aspect of human endeavors touching products and services in industry, government, and academia. When faults or vulnerabilities are exploited in this software, just as in proprietary software, the impacts can be far reaching. One detailed study on the depth to which OSS is entrenched in the critical infrastructures that service daily activities is detailed in the Linux Foundation's *Census II of Free and Open Source Software—Application Libraries* (Nagle et al., 2022).

The reuse of software has enabled faster fielding of systems since common components can be sourced externally. However, all software comes with vulnerabilities, and attackers have expanded their capabilities to exploit them in products that have broad use. A recent report by SecurityScorecard (Townsend, 2023) found that 98 percent of their sampled 230,000 organizations use third-party software components from organizations that have been breached within the prior two years.

Organizations shifting to cloud services to eliminate on-premises risk have frequently been surprised by supply chain risks inherited from service providers, which resulted from misconfigurations, unauthorized access, insecure application programming interfaces (APIs), etc. (Check Point Software Technologies Ltd., n.d.). To identify and manage this growing risk landscape, organizations must increase collaboration across the range of participants involved in the selection, installation, and monitoring of third-party software to identify and manage potential risks.

Earlier research addressed by the Software Engineering Institute (SEI) assembled practices critical to meeting this need in the *Acquisition Security Framework (ASF)* (Alberts et al., 2022). However, each organization has a unique technology environment, and there are no widely accepted measures for evaluating their accepted risk.

Diving into supply chain risk a little further, supply chain attacks initially appeared in third-party software that was either developed through custom contracts or purchased as commercial software. It is a bit easier to tailor a custom contract to protect against these risks, and the guidance available for supply chain risk management (SCRM) can go a long way towards developing a plan to add to a contract. Purchased software leaves the risk management to the customer to handle themselves. Establishing and implementing a SCRM plan is an expense that many organizations will not sign up for.

A recent European Union (EU) study (Papaphilippou et al., 2023) included 1,081 organizations in 27 member states. In their report, the researchers noted that only 47 percent of the surveyed organizations in the EU had an allocated budget for information and communication technology and operational technology (ICT/OT) for supply chain cybersecurity. Researchers further observed that 52 percent of the organizations surveyed had a rigid patching policy that included at least 80 percent of their assets. On the other hand, only 13.5 percent of the surveyed organizations had visibility into patching for fewer than half their assets, meaning that patching may be done by a third-party organization or not at all.

The situation is no better for users of OSS, which is the latest target of attackers in the software supply chain. For years, many considered OSS to be more secure because its code was visible and developed by "trusted" individuals. An early study (Hissam et al., 2002) provided insight into a widespread attack against both proprietary and OSS operating systems—the TearDrop and NewTear attacks—that were actually informed by OSS before vulnerability disclosure processes were widely used. Contrary to popular belief, OSS has never really been secure; whether a component of OSS is secure is in the eye of the beholder—be they altruistic or nefarious.

A Sonatype survey of 621 practitioners indicated that only 28 percent of their organizations became aware of new open source vulnerabilities within a day of disclosure, 39 percent discovered them within one to seven days, and 29 percent took more than a week to become aware of them (Sonatype Incorporated, 2023). The same survey found that 39 percent of the respondents' organizations took more than a week to mitigate vulnerabilities.

## Cost of an Attack

Intellias summarizes the estimated cost of supply chain attacks quite well (Fedorko, 2023). In 2023, the MOVEit vulnerability cost businesses over $9.9 billion, with more than 1,000 businesses and over 60 million individuals affected. Furthermore, the estimated cost from just seven high-profile supply chain attacks, starting with SolarWinds, was around $60 billion. This figure does not include the impact of government-imposed fines and legal actions related to privacy laws on both the affected organizations that rely on them. In this regard, there is little distinction between supply chain attacks resulting from compromised proprietary software and those resulting from compromised OSS.

Any successful supply chain attack can result in substantial financial loss, loss of reputation, lawsuits, and investigations. In recent articles (Birsan, 2021 & O'Neill, n.d.), it was revealed that the information of 1.6 million patients was compromised by a successful MOVEit hack. Even though the vulnerability was documented in May 2023, that particular hack was not discovered until October 2023. The patch to the vulnerability had not yet been applied.

In a report about the recovery from the successful attack on Metro-Goldwyn-Mayer (MGM) casinos (Jones, 2023), interviews with management indicated a loss of $100 million at its Las Vegas properties. The loss is covered by cyber insurance; however, the cost of that insurance has doubled or, in some cases, quadrupled in recent years. Of course, this loss estimate includes only the immediate loss of revenue and does not consider the result of class action lawsuits, some of which have already been filed. Typically, the organization plans to invest heavily in information technology (IT) **after** the successful attack.

Current efforts for improvement are largely focused on supply chain risk in general.

## Supply Chain Problem Space

Recently, a lot of attention has been paid to the problem of strengthening the software supply chain by focusing on software bill of material (SBOM) support. Although that attention is needed since there is little information available about the software specific to a product, it is not the be-all and end-all to securing supply chains. Yes, a well-defined and completely verifiable SBOM is a key step to understanding dependencies and tracking problems within those dependencies; however, it is not the only step. Just as important is knowing and understanding the practices employed by those who created the OSS itself—the OSS project members. For example, do the project members follow community standards of care when vetting contributions to the software code (e.g., peer reviews, code quality checks)? Having such transparency into the OSS project and the project's makeup is not unlike performing a financial health check or financial risk assessment of a commercial company for due diligence.

OSS consumers need to have the means and wherewithal to assess the products and projects they depend on. There are emerging OSS and proprietary tools to help provide insight into project health (e.g., the OSSF's Scorecard and MITRE's Hipcheck). There are also many standards, formats, and tools to help users maintain or create retroactive SBOMs that have been developed and promoted in recent years. However, to be effective, these tools must be conducive to the practices and processes employed by OSS consumers. Stakeholders in the entire OSS ecosystem and software industry in general still need novel tools, policies, and methodologies that detect and mitigate vulnerabilities found within OSS products including deviations from OSS-community-recommended practices in the products built to address new uses. By defining these tools, policies, and methodologies, the software community can establish greater confidence in its existing code base as well as assess and respond to new technologies, products, and vendors that provide great benefit.

There are also emerging OSS community initiatives being established. These are designed to provide additional insight into the practices employed by OSS projects. Two of note are the OSSF's Supply Chain Levels for Software Artifacts (OSSF's SLSA) and the Eclipse Foundation's Adoptium for Reproducible Builds. Each of these is an example of OSS community-led activities to improve confidence in OSS that are used by OSS consumers in their supply chains. In many ways, at least for security and supply chains, these examples are steps forward by the OSS community to "mature" OSS projects to get closer to secure software development and distribution.

## Software Risks and Challenges

Software development often lacks effective resistance to attack and controls for preventing tampering by malicious actors. Typically, software development project resources are insufficiently integrated and can be affected by cost, schedule, and compliance constraints. These characteristics leave operational environments with weaknesses and vulnerabilities that must be addressed later in the system's lifecycle. A lack of cost-schedule-risk balance can lead to shortcuts resulting in software that does not conform to secure coding and secure software development practices. There is a pressing need to implement more transparent and predictable mechanisms for ensuring that software functions securely and as intended.

No software is free of risks. Defects exist even in the highest quality software. For example, best-in-class code can have up to 600 defects per million lines of code (MLOC), while average quality code has around 6,000 defects per MLOC. Some of these defects are weaknesses that can lead to vulnerabilities. Research indicates that up to 5% of software defects are security vulnerabilities (Woody et al., 2014). As a result, best-in-class code can have up to 30 vulnerabilities per MLOC. For average quality code, the number of security vulnerabilities can be as high as 300 vulnerabilities per MLOC.

Reducing the number of security vulnerabilities in code is an important part of software development. Using secure coding practices, peer reviews, and code analysis tools are important ways to identify and correct known weaknesses and vulnerabilities in code. As shown in Figure 1, there are many activities that need to be addressed across the acquisition and development lifecycle to respond to defects and vulnerabilities as close to their point of origin as possible to keep costs for rework minimized.

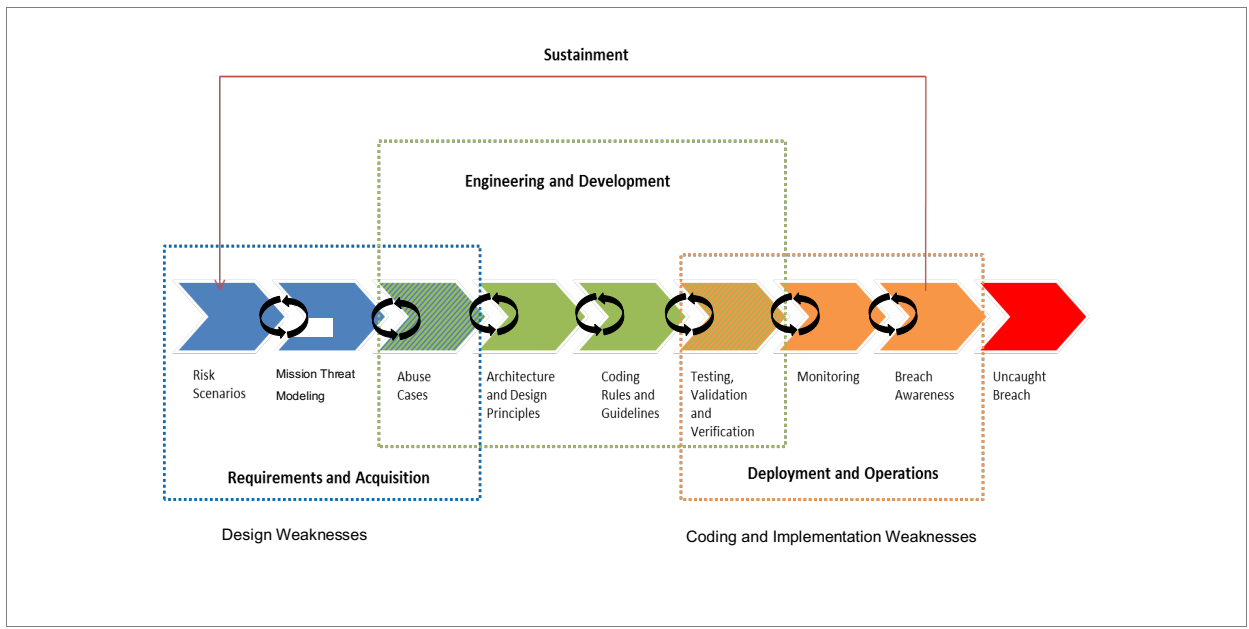

*Figure 1: Activities to Address Defect and Vulnerability Identification and Mitigation*

Third-party risk is a major challenge for organizations seeking to limit their exposure to cybersecurity risks from general outsourcing and supplier-provided software that is critical to system functions. There is often little transparency into the components, sources, and suppliers involved in a system's development and ongoing operation. SBOM information can help reduce the uncertainties that result from that lack of transparency. An essential aspect of addressing supplier risk is being able to access information about supplier inputs and their relative importance and then managing mitigations to reduce risk.

Research into the various types of vulnerabilities indicates that focusing on the most prevalent with greatest risk should be considered. Security vulnerabilities that are caused by defects with memory use are significant, highly common, and greatly impactful to the risk and damage to software, data, and systems. One recent study estimates that 65% of security vulnerabilities are caused by memory unsafety (Gaynor, 2020). Studies like this make addressing memory safety issues an appealing approach. Memory safety, however, is not without limitations. Known approaches cannot eliminate memory issues; they can only reduce them. And there are still other vulnerability categories that these mitigations do not address. Therefore, memory safety is not a panacea, but it can significantly reduce memory-related vulnerabilities, which would correspond to a significant reduction of risk. Like many other software security and quality mitigations, memory-related risks need to be measured and mapped to cybersecurity concerns to establish priority.

## State of the Practice of Measuring Software Assurance

Currently, it is possible to collect vast amounts of data related to cybersecurity in general. We can measure financial loss due to a successful attack and loss of confidence in a particular company as measured by the loss of customers or impacts on stock values. We can measure the elapsed time from the beginning of a successful attack until its discovery and the elapsed time from discovery until mitigation and recovery.

We can also measure specific product characteristics related to cybersecurity. Commercial companies, nonprofits, and government entities offer data-collection tools. What is not always clear is the cause-and-effect relationship between the data, vulnerabilities, and successful attacks. Also, much of the data collected reflects the results of an attack, whether attempted or successful. Data on earlier security lifecycle activities often reflects the development processes used. Although needed, it is not diligently collected, nor is it analyzed as thoroughly as in later points of the lifecycle.

As software engineers, we believe that improved software development practices and processes will result in a more robust and secure product. However, which specific practices and processes actually result in a more secure product? There can be quite a bit of elapsed time between the implementation of improved processes and practices and the subsequent deployment of the product. If the product is not successfully attacked, does it mean that it is more secure? Or does it just mean that it is a less interesting target from an attacker's perspective? Zahan concludes the following (Zahan, 2023):

> *Security metrics are a hard problem, especially in predicting vulnerabilities or assessing the effectiveness of counter measures [Cheng 2014, Scala 2019]. We should consider that the software security field has an inherently greater amount of unexpected variation.*

Consider the recently updated Cyber Security & Information Systems Information Analysis Center's (CSIAC's) *The DoD Cybersecurity Policy Chart* (Cybersecurity & Information Systems Information Analysis Center, n.d.). Although the authors have made a valiant effort to categorize and classify the subject policies, how can project managers be expected to sift through hundreds if not thousands of pages of documentation to find out what policy applies to them?

Certainly, government contractors have a profit motive that justifies meeting the cybersecurity policy requirements that apply to them, but do they know how to measure the cybersecurity risk of their products? And how would they know whether it has improved sufficiently? For OSS, when developers are not compensated, what would motivate them to meet these cybersecurity policy requirements? Why would they even care whether a particular organization—be it academic, industry, or government—is motivated to use their product?

## Currently Available Metrics

The SEI led a research effort to identify the metrics currently available within the lifecycle (Woody et al., 2019) that could be used to provide indicators that potential cybersecurity risk exists. From an acquisition lifecycle perspective, there are two critical information needs:

- Is the acquisition headed in the right direction as it is engineered and built (predictive)?
- Is the implementation sustaining an acceptable level of operational assurance (reactive)?

Each step in the acquisition and development lifecycle can produce useful information. However, to gather that information, the following must happen:

- The lifecycle's processes and practices must be structured to gather data in a form that can be analyzed.
- Appropriate measures must be created and enforced as each process or practice is performed.

Trends in current approaches show an increased reliance on tools to perform lifecycle steps and a limited consideration of how available data from tool use will be integrated across tasks for useful analysis.

For code, the work of Capers Jones in collecting and tracking defects to evaluate quality (Jones, 2014) provides an opportunity for setting expectations about code quality based on the level of expected defects. SEI research leveraged this data to project expected vulnerabilities based on defect rates (Woody et al., 2014). As development shifts further into Agile increments, many of which include third-party and open source components, different tools and definitions are applied to collecting defects so that the meaning of this metric in predicting risk becomes obscured.

Highly vulnerable components that are implemented using effective and well-managed zero trust principles can deliver acceptable operational risk. In a similar vein, well-constructed, high-quality components with weak interfaces can be highly prone to successful attacks. Operational context is critical to the level of risk exposure. A simple evaluation of each potential vulnerability using something like a Common Vulnerability Scoring System (CVSS) score can be extremely misleading since the score without the context has limited value in determining actual risk.

The lack of visibility into the processes and methods used to develop third-party software—particularly OSS—means that measures related to the processes used and the errors found prior to deployment, if they exist, do not add to the useful information about the product. This lack of visibility into product resilience as it relates to the process used to develop it means that we do not have a full picture of the risks, nor do we know whether the processes used to develop the product have been effective. It is difficult, if not impossible, to measure what is not visible.

Consider a recent blog post by Eric Goldstein from CISA (Goldstein, 2023) that announces a *Secure by Design Alert Series* and states the following:

> *Insecure technology products are not an issue of academic concern: they are directly harming critical infrastructure, small businesses, local communities, and American families.*

This, of course, is not new information. In addition to its implications for large companies producing software products, the same concerns apply to software in the supply chain **and** OSS. One might imagine that large companies will respond in some fashion, but how will the open source community respond? Are open source developers concerned with using specific languages, such as RUST, that are supportive of security goals? Are they concerned with using default settings that are inherently more secure? If so, how would consideration of these choices be communicated to customers?

Many ideas could apply to OSS, but there may be no agreement between the user and the creator of OSS. There are instances of organizations paying developers of OSS to provide software that is specific to their project. That software may or may not be covered by an agreement that would account for the cybersecurity risks. If there is no agreement between the user and the supplier, it leaves the user with the sole responsibility of defining performance measurements and determining whether or not they are being met. For the cybersecurity of OSS, by and large, the indicators are lagging (e.g., vulnerabilities, successful attacks) rather than leading.

When the processes used for product development are not visible, it is impossible to determine whether they have been effective in terms of thwarting future attacks. How can a process be improved when it is invisible and the measurement that should be taking place is also invisible? More to the point, how can we get leading indicators for OSS cybersecurity when the process is invisible?

## Assembling What We Can Know about OSS

The OSSF Scorecard is a tool that incorporates a set of metrics that can be applied to an OSS project. The idea is that project attributes that OSSF believes contribute to a more secure open source application are reported using a weighted approach that leads to a score.

From a metrics perspective, there are limitations to this approach:

1. The open source community is driving and evolving which items to measure and, therefore, build into the tool. Also, it is not clear how those factors were determined, whether the set of factors is complete, or what is intended for the long-term roadmap (i.e., insufficient transparency).
2. The relative importance of each factor is also built into the tool, which makes it difficult (but not impossible) to tailor the results to specific, custom, end-user needs (Open Source Software Foundation, n.d.).
3. Many of the items measured in the tool appear to be self-reported by developers versus validated by a third party, but this is a common "attribute" of open source projects.

Other tools, such as MITRE's Hipcheck, have the same limitations (MITRE Corporation, n.d.). For an OSSF project, it is possible to get a score using Scorecard and scores for the individual dependency projects, but questions arise from this approach. How do those individual scores roll up into the overall score? Do you pick the lowest score across all the dependencies? Or do you apply some sort of weighted average of scores? This area needs exploration and elaboration.

Furthermore, a recent research paper (Zahan et al., 2023) described cases where open source projects that were scored highly by Scorecard might, in fact, produce packages that have more reported vulnerabilities. From a research perspective, it is unknown whether this occurs because (1) the application received more reviews and therefore more vulnerabilities were identified, or (2) attacks on a popular application exposed it to more vulnerabilities. Needless to say, Zahan's results are useful only for those open source projects evaluated by Scorecard, which is applied exclusively to GitHub, and those are only a fraction of the total number of open source applications available. All of these issues indicate that further study is needed.

Metrics by themselves are of limited value. It is the assembled picture of metrics weighed against expectations that can be valuable to decision makers. A simple three-step process can be applied to generate an initial perspective and flag areas of concern for further consideration:

- Measure and baseline what you have, especially OSS.
- Assess how you are vulnerable and identify an improvement path.

- Integrate measurement and monitoring throughout the lifecycle.

An example, as shown in Figure 2, can provide a useful starting point even if information is limited. Known information related to the project, product, available protection, and applied policies provide a useful range of sources. Mapping that information to a series of potential factors can raise red flags to show where security risk may already exist. Evaluating each of the risks against a range of acceptability ratings can identify areas of concern that will require mitigation.

| Identified Criteria | Project | Product | Protection | Policy |
|---|---|---|---|---|
| Long-Term Support | Forked Project | | | No Security Policy |
| Dependencies | 74 Abandoned Dependencies | No Update Tools | | |
| Security | | 4 Unfixed Critical Vulnerabilities | Workflow with Excessive Permissions | |
| Integrity | | No Fuzz Testing | 30 Unreviewed Change Sets | |
| Malicious Actors | Commit ID Known Malicious | | | |
| Suitability | | | | 12 Restrictive Licenses |

Realm of Observable Facts of OSS Projects and Products

Red Flag

*Figure 2: Mapping Observable Information to Areas of Concern*

The steps described in Figure 2 can be listed as follows:

- Review data available.
- Identify useful criteria.
- Extract key data.
- Map to acceptable criteria.
- Evaluate red flags.
- Identify appropriate mitigations.
- Confirm supportability.

When obtaining information, there will be challenges about the processes used to create and maintain OSS since this is information typically not captured in available repositories. When issues arise, there are limited options at best for mandating repairs.

## Future Considerations

At a March 2024 OSS Summit hosted by CISA, participants from several open source communities gathered to discuss a path forward to improve the security and potentially reduce the risk of using OSS.

Suppliers should be proactive. Yet leadership across the supply chain continues to underinvest in software assurance, especially early in the lifecycle. This lack of investment leads to design decisions that lock in weaknesses because there are no means to characterize and measure the risk they are accepting. Suppliers rush to deliver new features to motivate buyers at the expense of analyzing the code to remove potential vulnerabilities, and buyers have limited means to evaluate the risk in the products they acquire.

Even if a supplier addresses an identified vulnerability quickly and issues a patch, it is incumbent on the users of that software to apply the fix. Software supply chains are many levels deep, and too frequently the patches apply to products buried deep within the chain. Each layer must apply the patch and send an update up the chain. This can be a slow and faulty process since knowing where each specific product has been used is limited for those higher in the chain. Recent mandates to create SBOMs (United States White House Executive Office of the President, 2022) support improving that visibility, but the fix still needs to be addressed by each of the many layers that contain the vulnerability.

A better future is needed, where the tools we have available to assess software packages for supply chain risk perform comparable analyses in comparable ways, where the results of analyses are clear and can be tied to policy needs, and where the recommendations made by analyzers can be audited to provide clarity to software maintainers and security teams alike. Achieving this future means going beyond simple observation analysis and establishing a process to define standard metrics for supply chain security risk in software.

## Reference Documents Used to Prepare this Paper

The following documents were used to prepare this paper:

- The Measurement Challenges in Software Assurance and Supply Chain Risk Management. (Mead et al., 2023)
- The SEI's response (Tucker et al., 2023) to the Office of the National Cyber Director's Request for Information on Open-Source Software Security: Areas of Long-Term Focus and Prioritization (United States Office of the National Cyber Director, 2023).

## Author Bios

**Dr. Carol Woody** is a principal researcher in the CERT Division of the Software Engineering Institute. She focuses on cybersecurity engineering for building capabilities and competencies to measure, manage, and sustain cybersecurity and software assurance for highly complex software-reliant systems and systems of systems. She has been a member of the CERT technical staff for over 20 years. Dr. Woody coauthored the book *Cyber Security Engineering: A Practical Approach for Systems and Software Assurance,* which was published as part of the SEI Series in Software Engineering. The CERT Cybersecurity Engineering and Software Assurance Professional Certificate, a self-paced online training program, is based on research she led.

**Dr. Nancy Mead** is a Fellow at the Software Engineering Institute and Adjunct Professor of Software Engineering at Carnegie Mellon University. Her research areas are software and security requirements engineering and software engineering and software assurance curricula. The Nancy Mead Award for Excellence in Software Engineering Education & Training is named for her. She has more than 200 publications and invited presentations. Her awards and honors include Life Fellow of the IEEE, Distinguished Member of the ACM; IEEE TCSE Distinguished Educator; IEEE TCSE Executive Board Member; Parnas Fellow at Lero, the Irish Software Research Center; Speaker in the IEEE Distinguished Visitor Program. She has a BA, MS, and PhD in mathematics from NYU.

**Scott A. Hissam** is a senior member of the technical staff for Carnegie Mellon University's Software Engineering Institute where he conducts research on component-based software engineering, open source software, and software assurance through his interactions with government, industry, and academic ventures. He is a founding member of the International Federation for Information Processing (IFIP) Working Group 2.13 on Open Source Software and co-organizer of a number of its annual conferences. His publications include two books (*Building Systems from Commercial Components* and *Perspectives on Free and Open Source Software*), papers published in international journals, and numerous technical reports. Most recently, Mr. Hissam has been focusing his time on engaging various open source software communities, open source program offices, and government program offices in best practices and their application in software-and cyber-supply chain risk management in a DoD environment.

## Legal Markings